\documentclass[%
 reprint,
nofootinbib,
 amsmath,amssymb,
 aps,
pra,
]{revtex4-1}

\usepackage{graphicx}% Include figure files
\usepackage{dcolumn}% Align table columns on decimal point
\usepackage{bm}% bold math
\usepackage[utf8]{inputenc}
\usepackage[T1]{fontenc}
\usepackage{mathptmx}
\usepackage{xcolor}

\begin{document}

\preprint{APS/123-QED}

\title[]{Bounds on bipartite entanglement from fixed marginals}% Force line breaks with \\

\author{Giuseppe Baio\textsuperscript{1}}
\author{Dariusz Chru\'{s}ci\'{n}ski\textsuperscript{2}}%
\author{Paweł Horodecki\textsuperscript{3,4}}
\author{Antonino Messina\textsuperscript{5}\vspace*{0.2cm}}
\author{Gniewomir Sarbicki\textsuperscript{2}}%

\email{giuseppe.baio@strath.ac.uk}
\affiliation{
\textsuperscript{1}\textit{SUPA and Department of Physics, University of Strathclyde, Glasgow G4 0NG, Scotland, U.K.}}
\affiliation{\textsuperscript{2}\textit{Nicolaus Copernicus University,  Grudzi\k{a}dzka 5/7, 87–100 Toru\'{n}, Poland}}
\affiliation{\textsuperscript{3}\textit{\small International Centre for Theory of Quantum Technologies,
University of Gdansk, Wita Stwosza 63, 80-308 Gdansk, Poland }}
\affiliation{\textsuperscript{4}\textit{\small
Faculty of Applied Physics and Mathematics, National Quantum Information Centre,Gdansk
University of Technology, Gabriela Narutowicza 11/12, 80-233 Gdansk, Poland }}
%Faculty of Applied Physics and Mathematics, Gdańsk University of Technology,
%Gabriela Narutowicza 11/12, 80-233 Gdańsk, Poland
%and National Quantum Information Center of Gdańsk,
%Władysława Andersa 27, 81-824 Sopot, Poland}}
\affiliation{\textsuperscript{5}\textit{\small Dipartimento di Matematica e Informatica, Università degli Studi di Palermo, Italy and I.N.F.N., Sezione di Catania, Italy}}

\date{\today}% It is always \today, today,

\begin{abstract}
We discuss the problem of characterizing upper bounds on entanglement in a bipartite quantum system when only the reduced density matrices (marginals) are known. In particular, starting from the known two-qubit case, we propose a family of candidates for maximally entangled mixed states with respect to fixed marginals for two qudits. Interestingly, it turns out such states are always quasidistillable. Moreover, they are extremal in the convex set of two qudit states with fixed marginals. Our observations are supported by numerical analysis.

%\begin{description}
%\end{description}
\end{abstract}

\pacs{Valid PACS appear here}% PACS, the Physics and Astronomy
\maketitle

%\tableofcontents

\section{\label{sec:level1}Introduction}

The preparation of a quantum system in a certain state  is regarded as a central target in several contexts and if the system is multipartite, the possible \textit{entanglement} among subsystems is an useful resource for quantum information processing and quantum communication \cite{NC}. Suitable criteria to characterize or quantify entanglement are then of primary importance \cite{horodecki}.
For pure bipartite states $\rho_{AB}=|\Psi_{AB}\rangle\langle\Psi_{AB}|$ with $|\Psi_{AB}\rangle\ \in  \mathcal{H}_A \otimes \mathcal{H}_B$, the Von Neumann entropy of any of the two reduced density matrices or \textit{marginals} reads:

\begin{equation}
E(\Psi_{AB}) = S(\rho_{A})=-\textrm{Tr}(\rho_{A} \log \rho_{A})
\label{vonneum}
\end{equation}
where $\rho_{A}=\textrm{Tr}_{B}|\Psi_{AB}\rangle\langle\Psi_{AB}|$. For mixed states the situation is much more complicated
and the simple formula is replaced by the convex roof construction leading to the well known entanglement of formation EOF defined by \cite{EOF}

\begin{equation}\label{}
  {\rm EOF}(\rho_{AB}) = \min_{p_k,\Psi_k} \sum_k p_k E(\Psi_k)
\end{equation}

where the minimum is performed over all decompositions $\rho_{AB} = \sum_k p_k|\Psi_k\rangle\langle\Psi_k|$. For the two qubit case EOF can be reduced to the celebrated Wootters {concurrence}:

\begin{equation}
C(\rho_{AB})\equiv \max\{0,\alpha_{1}-\alpha_{2}-\alpha_{3}-\alpha_{4}\}
\label{concdef}
\end{equation}
where $\{\alpha_{i}\}$ are the square roots of the four eigenvalues of the matrix  $\rho_{AB}(\sigma_{y}\otimes\sigma_{y})\rho_{AB}^{*}(\sigma_{y}\otimes\sigma_{y})$ taken in decreasing order \cite{hill} (for an introduction to entanglement measures see for example the review \cite{Virmani}).

A simple way to characterize mixed bipartite entanglement is based on the Peres-Horodecki criterion, also known as \textit{PPT condition} \cite{peres1,horodecki2}: if a state $\rho_{AB}$ is separable then its \textit{partial transposition} $\rho_{AB}^{\tau}=\left(\mathbb{I}\otimes\tau\right)\rho_{AB}$ is necessarily positive semidefinite. Such condition becomes necessary and sufficient only for the two-qubits and qubit-qutrit cases \cite{horodecki3}. A  measure of entanglement called \textit{negativity} can be defined as follows:

\begin{equation} \label{negdef}
N(\rho_{AB}) \equiv \frac{1}{2} \left(\left\lVert \rho_{AB}^{\tau}\right\rVert_1-1\right)
\end{equation}
where $\left\lVert \rho_{AB}^{\tau}\right\rVert_1$ is the trace norm of $\rho_{AB}^{\tau}$. Such definition provides a convex function which is non increasing under local operation and classical communication \cite{vidalwerner, plenio}.

A relevant feature of mixed bipartite states is the relation between entanglement and purity \cite{gurvits}. In particular, for a given purity $P=\textrm{Tr}(\rho_{AB}^2)$, one may ask which state of the same purity displays maximal entanglement \cite{zyczkowski}. The concept of maximally entangled mixed state (MEMS) for two-qubits was introduced by Ishizaka and Hiroshima as  states such that any entanglement measure cannot be increased by any global unitary \cite{MEMS1}. They proposed a family of optimal states that was also supported by Munro \textit{et al.} \cite{MEMS2}.
Such family is recovered by means of the transformation maximizing the entanglement in the spectrum constrained analogue problem, found by Verstraete \textit{et al.} \cite{MEMS3}. Despite recent numerical efforts, no analytical results are  available for higher dimensional cases \cite{hedemann,mendonca}.

In this paper we analyze a similar problem. We ask what is the maximal entanglement achievable by a bipartite system with fixed marginal states $\rho_{A}$ and $\rho_{B}$.
Such an assumption of fixed marginals is known to introduce constraints on the spectrum of the joint state $\rho_{AB}$ in the form of linear inequalities, as shown by Klyachko  \cite{klyachko1,klyachko2}.
However, such constraints do not directly tell about possible correlations among the subsystems. %restrictions on the spectrum are restrictions on the purity and then on the entanglement%
Therefore, focusing on bipartite entanglement in such scenario, we investigate MEMS with respect to fixed marginals which provide the upper bound on entanglement stemming from local information only.

The paper is organised as follows. In section \ref{sec:level2} we review the known results for two qubit states, including a characterization of the optimal states as extremal points of the convex set with fixed marginals, originally discussed in \cite{parthasarathy, rudolph}. In section \ref{sec:level3} we present a family of candidate MEMS with respect to fixed marginals, supported by an insightful physical interpretation from the point of view of entanglement distillation. Finally, in section \ref{sec:level4}, we present numerical studies comparing our candidate states with the results of numerical optimization for the case of two-qutrits which support our conjecture.

\section{\label{sec:level2}Known results}

The problem of characterizing mixed bipartite entanglement of states with fixed marginal properties was first introduced in \cite{adesso}. In particular, a special class of two-qubits states under scrutiny there was denoted as maximally entangled \textit{marginally} mixed states (MEMMS), i.e. MEMS with respect to certain local purities. Clearly, only in the two-qubit case, a given value of both  $P_{A}=\textrm{Tr}(\rho_{A}^2)$ and $P_{B}=\textrm{Tr}(\rho_{B}^2)$ uniquely determines the local spectra. Throughout the work, we assume instead complete knowledge of the marginal states.

\subsection{Two-Qubits case}

Let us start our analysis in a pedagogical fashion and introduce a suitable representation of states with fixed marginals. This is described only in the two-qubit case %($\mathcal{H}=\mathbb{C}^{2}\otimes\mathbb{C}^{2}$)
but its generalization to arbitrary high dimensions is straightforward. Let $\rho_{A}$ and $\rho_{B}$ be two qubit states. We fix local bases such that the states of the two subsystems are given in diagonal form:

\begin{equation}
\rho_{A}=\textrm{diag}\{1-\lambda_{A},\lambda_{A}\},\quad\rho_{B}=\textrm{diag}\{1-\lambda_{B},\lambda_{B}\}
\label{margdef}
\end{equation}
with the lowest eigenvalues such that $\lambda_{A},\lambda_{B}\in\left[0,\frac{1}{2}\right]$. Assuming the ordering $\lambda_{A}\geq\lambda_{B}$, any joint state $\rho_{AB}$ with marginals (\ref{margdef}) can be represented  as follows:

\begin{equation}
\rho_{AB}=\rho_{A}\otimes\rho_{B}+\Delta
\label{corre}
\end{equation}
where $\Delta$ is such that $\textrm{Tr}_{A}\Delta=\textrm{Tr}_{B}\Delta=0$ and it contains all possible correlations, quantum and classical, admitted by the two subsystems compatible with fixed marginals $\rho_{A}$ and $\rho_{B}$. It is easy to see that the most general two-qubit matrix form of (\ref{corre}) is the following \cite{baio}:

\begin{equation}\begin{split}
\rho_{AB}=\rho_{A}\otimes\rho_{B}+
\left(\begin{array}{cc|cc}
\epsilon&\Delta_{12}&\Delta_{13}&\Delta_{14}\\
&-\epsilon&\Delta_{23}&-\Delta_{13}\\
\hline
&&-\epsilon&-\Delta_{12}\\
\textrm{(c.c)}&&&\epsilon\end{array}\right)
\label{form}
\end{split}\end{equation}
where one has to choose the entries of $\Delta$ such that $\rho_{AB}\geq 0$.
Let us observe that, in order to obtain a non-negative diagonal elements, one finds

\begin{equation}
-\lambda_{A}\lambda_{B}\leq\epsilon\leq\lambda_{B}(1-\lambda_{A}).
\label{rangeepsilon}
\end{equation}
%Once a value for $\epsilon$ is chosen, a simple approach to ensure $\rho_{AB}\geq 0$ is to search a \textit{factorizable} form, i.e. $\rho_{AB}=LL^{\dag}$, where $L$ is any complex matrix $\in\mathcal{M}_{m,n}(\mathbb{C})$. For example, this can be achieved by means of Cholesky algorithm.  In this case $L$ is lower triangular and one derives conditions such that the decomposition exists \cite{matrix}. %
Two-qubit MEMMS states are thus achieved maximizing concurrence or negativity of states in the form (\ref{form}). However, in this case it is sufficient to consider the subclass of  X-states only (non zero diagonal and anti-diagonal) since it includes also the two-qubit MEMS \cite{MEMS1,MEMS2,MEMS3,quasi3}. X-states are common in quantum information theory because of their sparse structure, allowing for many analytic computations \cite{mendonca2}. Important families of two-qubit states such as Bell, Werner or isotropic states are within this class.   Hence we consider:

\begin{equation}\begin{split}
\rho_{AB}=\rho_{A}\otimes\rho_{B}+
\left(\begin{array}{cc|cc}
\epsilon&\cdot&\cdot&\Delta_{14}\\
&-\epsilon&\Delta_{23}&\cdot\\
\hline
&&-\epsilon&\cdot\\
\textrm{(c.c)}&&&\epsilon\end{array}\right).
\label{class}
\end{split}\end{equation}
Such simple structure yields the following concurrence:

\begin{equation}\begin{split}
C(\rho_{AB})&=2\max\{0,\left|\Delta_{23}\right|-\sqrt{\rho_{11}\rho_{44}},\left|\Delta_{14}\right|-\sqrt{\rho_{22}\rho_{33}}\}
\label{concX}
\end{split}\end{equation}
where $\{\rho_{ij}\}_{i,j=1,\dots,4}$ are matrix elements of $\rho_{AB}$ \cite{hill}. It is useful, according to (\ref{rangeepsilon}), to parameterize $\epsilon$ via $
\epsilon=s\lambda_{B}-\lambda_{A}\lambda_{B}$, where $s\in\left[0,1\right]$. Finally, positivity of $\rho_{AB}$ is simply controlled by the following inequalities for a given $s$:

\begin{equation}\begin{split}
\left|\Delta_{23}\right|^{2}\leq \lambda_{B}(1-s)(\lambda_{A}-\lambda_{B}s),\\
\left|\Delta_{14}\right|^{2}\leq s\lambda_{B}(1-\lambda_{A}-\lambda_{B}+\lambda_{B}s).\\
\label{ineqpos}
\end{split}\end{equation}
Due to the simplicity of (\ref{concX}), one can independently maximize both RHS of (\ref{ineqpos}) and observe that the maximum is reached when $s=1, |\Delta_{23}|=0,|\Delta_{14}|=\sqrt{(1-\lambda_{A})\lambda_{B}}$, giving rise to the following state:

%\begin{figure}\begin{center}
%\includegraphics[scale=0.44]{defini.pdf}
%\caption{Example 2 - Plot to be improved}
%\label{Delta23}
%\end{center}\end{figure}

\begin{equation}
\tilde{\rho}_{AB}=
\left(\begin{array}{cc|cc}
1-\lambda_{A}&\cdot&\cdot&\sqrt{(1-\lambda_{A})\lambda_{B}}\\
\cdot&0&\cdot&\cdot\\
\hline
\cdot&\cdot&\lambda_{A}-\lambda_{B}&\cdot\\
\sqrt{(1-\lambda_{A})\lambda_{B}}&\cdot&\cdot&\lambda_{B}\end{array}\right)
\label{MEMMS}
\end{equation}
with negativity given by:

$$ N(\tilde{\rho}_{AB})= \frac{1}{2} \left(\lambda_A-\lambda_B-\sqrt{(\lambda_A-\lambda_B)^2+4\lambda_B(1-\lambda_A)}\right). $$

\begin{figure}[h!]
\hspace*{-0.35cm}\includegraphics[scale=0.645]{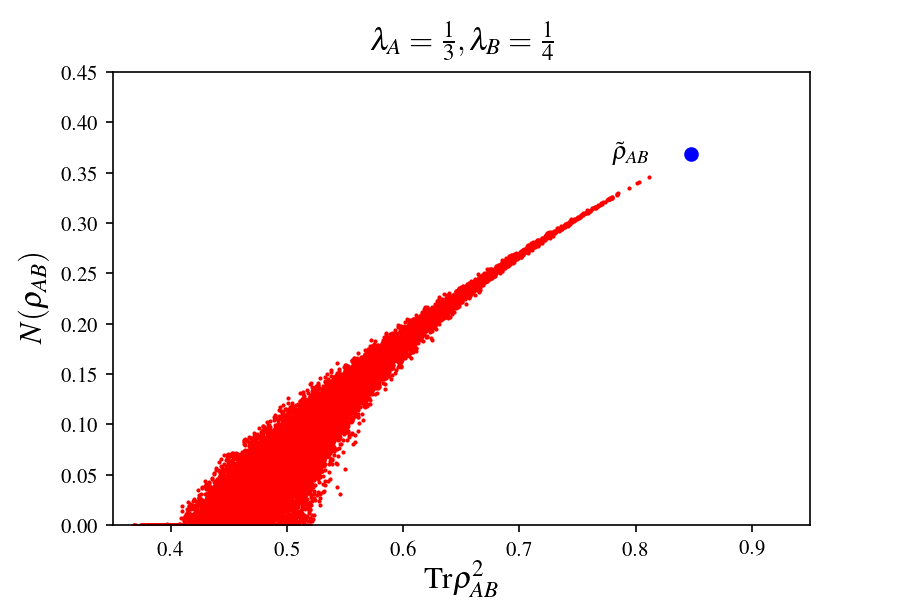}
\caption{N-P plot with 20000 randomly generated two-qubit states with  marginals $\lambda_{A}=\frac{1}{3}$ and $\lambda_{B}=\frac{1}{4}$. Note that both purity and negativity are maximised by the same state $\tilde{\rho}_{AB}$ (\ref{structure}).}
\label{random}
\end{figure}

This represents the upper bound for a two-qubit system with arbitrarily fixed marginals, in accordance with \cite{adesso}.
Interestingly,  $\tilde{\rho}_{AB}$ can be written as follows:

\begin{equation}
\tilde{\rho}_{AB}=(1-\eta)|\Psi_{\textrm{mc}}\rangle\langle\Psi_{\textrm{mc}}|+ \eta|10\rangle\langle 10|
\label{structure}
\end{equation}
where $\{|0\rangle, |1\rangle\}$ is the computational basis in $\mathbb{C}^2$, $\eta=\lambda_{A}-\lambda_{B}$ and $|\Psi_{\textrm{mc}}\rangle\langle\Psi_{\textrm{mc}}|$ is a \textit{maximally correlated} rank-1 projector. Recall, that a state $\sigma_{\textrm{mc}}$ maximally correlated (or Schmidt-correlated) in the computational basis in $\mathbb{C}^{d}\otimes\mathbb{C}^{d}$ reads \cite{rains}:

\begin{equation}
\sigma_{\textrm{mc}} = \sum^{d-1}_{i,j=0}\alpha_{ij} |ii\rangle\langle jj|.
\label{maxcor}
\end{equation}
Moreover, the state (\ref{maxcor}) has all its (at most d) eigenvectors in the form $|\Psi_{k}\rangle=\frac{1}{\sqrt{d}}\sum_{k}\lambda^{(k)}_{i}|ii\rangle$.
In order to provide a simple visual representation that (\ref{MEMMS}) is the optimal state we construct a negativity vs. global purity plot (N-P), shown in Fig. \ref{random}. This allows us to compare the negativity of $\tilde{\rho}_{AB}$ with that of a set of randomly generated states from (\ref{form}).
In what follows we briefly recall a further characterization of the optimal state as an extremal point of a convex set. The motivation is simple: negativity is a convex function and the set of states with fixed marginals is also convex, hence the  maximum must be attained by an extremal point \cite{convex1}.

\subsection{Optimal states as extremal points}

Let us denote with $C(\rho_A,\rho_B)$ the convex set of two-qudit states with fixed marginals $\rho_A$ and $\rho_B$. The characterization of the extremal points of $C(\rho_A,\rho_B)$ was provided first by Parthasarathy in \cite{parthasarathy}. Here we follow instead the approach by Rudolph, based on the duality between positive operators and completely positive (CP) maps \cite{rudolph}.
We recall that a map $\Lambda$ is CP iff the map $\textrm{id}_{k}\otimes\Lambda$ is positive $\forall\, k \in \mathbb{N}_{+}$ where $\textrm{id}_{k}$ denotes the identity map. A powerful tool providing a duality between states and CP maps is given by the  Choi-Jamio\l{}kowski isomorphism \cite{Jamio}. For each CP map $\Lambda$, one can assign a legitimate density matrix $\rho_{\Lambda}$:

\begin{equation}
\rho_{\Lambda} = [\textrm{id}_{d}\otimes\Lambda](P^+_{d}), \quad P^+_{d} = \frac{1}{d} \sum^{d}_{i,j=1}|ii\rangle\langle jj|
\label{directiso}
\end{equation}
where $P^+_{d}$ is a maximally entangled projector and ${\rm id}_d$ is an identity map.  Since the above duality is bijective, one has the following inverse CP map for any state $\rho$:

\begin{equation}
\Lambda_{\rho}[\sigma] = \textrm{Tr}_{2}[\textrm{id}_{d}\otimes\sigma^{T}\rho].
\label{inverse}
\end{equation}
This allows us to describe the convex structure of the set of states with fixed marginals at the level of the corresponding maps. In particular, one can exploit the known characterization of extremal maps in terms of their Kraus representation of $\Lambda_{\rho}[\sigma] = \sum_{\alpha}K_{\alpha}\sigma K^{\dagger}_{\alpha}$, namely that $\rho_{\Lambda}$ is extremal iff $\{K^{\dagger}_{\alpha}K_{\beta}\}_{\alpha,\beta=1,\dots,d^2}
\label{extcond}$ is a linearly independent set of matrices \cite{choi}. Moreover, the constraint of fixed marginals $\rho_A$ and $\rho_B$ are expressed via eq. (\ref{inverse}):

\begin{equation}\begin{split}
\frac{1}{d}\Lambda_{\rho}(\mathbb{I}_{d}) = \frac{1}{d} \sum_{\alpha}K_\alpha K^{\dagger}_\alpha = \rho_A\\
\frac{1}{d}\Lambda^*_{\rho}(\mathbb{I}_{d}) = \frac{1}{d} \sum_{\alpha}K^{\dagger}_\alpha K_\alpha = \rho_B
\label{extcond1}
\end{split}\end{equation}
where $\Lambda^*$ denotes the canonical dual. Thus, the extremality condition amounts at proving that the set:

\begin{equation}
\{K^{\dagger}_{\alpha}K_{\beta}\oplus K_{\beta}K^{\dagger}_{\alpha}\}_{\alpha,\beta=1,\dots,d^2}
\label{extcond2}
\end{equation}
is linearly independent, i.e. the two sets $\{K^{\dagger}_{\alpha}K_{\beta}\}_{\alpha,\beta=1,\dots,d^2}$  and $\{K_{\beta}K^{\dagger}_{\alpha}\}_{\alpha,\beta=1,\dots,d^2}$ are \textit{jointly} linearly independent \cite{landau}. As an example, we have that the only extremal two-qubit state for $C(\frac{1}{2}\mathbb{I}_{2},\frac{1}{2}\mathbb{I}_{2})$ is the maximally entangled projector $P^+_{2}$ \cite{parthasarathy, rudolph}. The criterion given by conditions (\ref{extcond1}) and (\ref{extcond2}) can be applied to our case in order to construct examples of extremal points in $C(\rho_A,\rho_B)$. Note that the optimal rank-2 state (\ref{MEMMS}) is retrieved by means of the following Kraus operators:

\begin{equation}\begin{split}
&K_1 = \left(
\begin{array}{cc}
 0 & 0 \\
 \sqrt{\lambda_A-\lambda_B} & 0 \\
\end{array}
\right),\\
&K_2 =\left(\begin{array}{cc}
 \sqrt{1-\lambda_A} & 0 \\
 0 & \sqrt{\lambda_B} \\
\end{array}
\right).
\label{kraus12}
\end{split}\end{equation}
One can easily check that the fixed marginals and extremality conditions hold (cf. Appendix A). Moreover, defining $e_{ij}=|i\rangle\langle j|$, the corresponding rank-2 extremal, given by (\ref{directiso}), reads:

\begin{equation}
\rho_{\Lambda}=\frac{1}{2}\sum^{2}_{i,j=1}\sum^{2}_{\alpha=1}e_{ij}\otimes K_{\alpha}\,e_{ij}\,K^{\dagger}_\alpha
\end{equation}

and coincides with the optimal state (\ref{MEMMS}). The parametrization of the class of extremal states for arbitrarily given marginals in higher dimensions ($\mathbb{C}^d\otimes\mathbb{C}^d, \,d\geq 2$) is out of the aim of this work and will not be discussed here. Nevertheless, we will adopt in the next section the extremality condition as further check on the candidate MEMS with respect to fixed marginals. One can find the following necessary condition for extremal points in $C(\rho_A,\rho_B)$ \cite{parthasarathy}:

\begin{equation}\label{}
  \textrm{rank}(\rho)\leq\sqrt{2d^2-1} .
\end{equation}
This observation turns out to be useful for the numerical studies discussed later in section \ref{sec:level4}.

\section{\label{sec:level3}Higher Dimensions}

In this section we discuss the properties of a family of states within which we identify candidates for two-qutrit MEMS with respect to marginals. A crucial observation is that all candidate states are \textit{quasidistillable}, i.e.  states for which a singlet fraction arbitrarily close to unity can be obtained in the distillation process \cite{quasi}. A connection between MEMS and quasidistillable states was highlighted previously in \cite{quasi2, quasi3}.

\subsection{A family of candidates}

As an attempt to directly generalize the two-qubits results, we focus on the extension of the form (\ref{structure}) to higher dimensions $\mathbb{C}^d\otimes\mathbb{C}^d$, $d\geq2$, namely:

\begin{equation}
\tilde{\rho}=(1-\eta)\sigma_{\textrm{mc}}+\sum_{i\neq j} p_{ij}|ij\rangle\langle i j|
\label{familydef}
\end{equation}
where $\eta=\sum_{i\neq j}p_{ij}$ and $\sigma_\textrm{mc}$ indicates a maximally correlated state of the form (\ref{maxcor}). Note that replacing $\sigma_{\textrm{mc}}$ with $P^{+}_{d}$, one obtains a possible generalization of isotropic states \cite{dariusz1}. Furthermore, the family defined by eq. (\ref{familydef}) belongs to a wider class known as \textit{circulant} states which reduces to X-states in $\mathbb{C}^2\otimes\mathbb{C}^2$  \cite{dariusz2}. For the rest of the work, we examine the two-qutrit case for which the matrix structure of (\ref{familydef}) reads:

\small
\begin{equation}
\tilde{\rho}=
\left(\begin{array}{ccc|ccc|ccc}
\rho_{11}&\cdot&\cdot&\cdot&\Delta_{15}&\cdot&\cdot&\cdot&\Delta_{19}\\
&\rho_{22}&\cdot&\cdot&\cdot&\cdot&\cdot&\cdot&\cdot\\
&&\rho_{33}&\cdot&\cdot&\cdot&\cdot&\cdot\\
\hline
&&&\rho_{44}&\cdot&\cdot&\cdot&\cdot&\cdot\\
&&&&\rho_{55}&\cdot&\cdot&\cdot&\Delta_{59}\\
&&&&&\rho_{66}&\cdot&\cdot&\cdot\\
\hline
&&&&&&\rho_{77}&\cdot&\cdot\\
&&&&&&&\rho_{88}&\cdot\\
(\textrm{c.c.})&&&&&&&&\rho_{99}\\
\end{array}\right)
\label{circstruct}
\end{equation}
\normalsize

and satisfies  $\textrm{Tr}\tilde{\rho}=1$, positivity condition and compatibility with the following marginals:

\begin{equation}\begin{split}
&\rho_{A}=\textrm{diag}\left\{
1-\lambda_{1}-\lambda_{2},\lambda_{1},\lambda_{2}
\right\},\\
&\rho_{B}=\textrm{diag}\left\{1-\mu_{1}-\mu_{2},\mu_{1},\mu_{2}\right\},
\end{split}\end{equation}
where $\lambda_{1}\geq\lambda_{2}$, $\mu_{1}\geq\mu_{2}$ correspond to the decreasingly ordered local eigenvalues and, thus, $\lambda_{2},\mu_{2}\leq\frac{1}{3}$. Without loss of generality we also assume $\lambda_{1}+\lambda_{2}\geq\mu_{1}+\mu_{2}$. The negativity of (\ref{circstruct}) is simply given by:

\begin{equation}
N(\tilde{\rho})=
\frac{1}{2}\left[(\left|A\right|-A)+(\left|B\right|-B)+(\left|C\right|-C)\right]
\end{equation}
where:

\vspace*{-0.5cm}
\begin{equation}\begin{split}
A=\frac{1}{2} \left(\rho_{22}+\rho_{44}-\sqrt{4 \left|\Delta_{15}\right|^{2}+\left(\rho_{22}-\rho_{44}\right){}^2}\right),\\
B=\frac{1}{2} \left(\rho_{33}+\rho_{77}-\sqrt{4 \left|\Delta_{19}\right|^{2}+\left(\rho_{33}-\rho_{77}\right){}^2}\right),\\
C=\frac{1}{2} \left(\rho_{66}+\rho_{88}-\sqrt{4 \left|\Delta_{59}\right|^{2}+\left(\rho_{66}-\rho_{88}\right)^2}\right).
\label{eigenvaluesPT}
\end{split}\end{equation}
Note that if at least one of the diagonal elements in each term of (\ref{eigenvaluesPT}) is zero, we already reach the maximum number of negative eigenvalues of the partial transpose. Moreover, $N(\tilde{\rho})$ increases monotonically with $|\Delta_{ij}|$ and thus it is favourable to have the maximum number of zeros (four) in the diagonal which can always be chosen  independently in (\ref{eigenvaluesPT}). Maximum negativity within our family is then attained by the following three states:

\begin{equation}\begin{split}
&\tilde{\rho}_{AB}^{(1)}=\left(1-p_{10}-p_{20}\right)\,|\Psi^{(1)}_{\textrm{mc}}\rangle\langle\Psi^{(1)}_{\textrm{mc}}|+p_{10}\,|10\rangle\langle 10| + p_{20}\,|20\rangle\langle 20|\\
&p_{10} = \lambda_1 -\mu_1,\,  p_{20} = \lambda_2 -\mu_2
\label{candidate1}
\end{split}\end{equation}
valid when $\lambda_{1}>\mu_{1}$ and $\lambda_{2}>\mu_{2}$,

\begin{equation}\begin{split}
&\tilde{\rho}_{AB}^{(2)}=\left(1-p_{10}-p_{12}\right)\,|\Psi^{(2)}_{\textrm{mc}}\rangle\langle\Psi^{(2)}_{\textrm{mc}}|+p_{10}\,|10\rangle\langle 10| + p_{12}\,|12\rangle\langle 12|\\
& p_{10} = \lambda_1 + \lambda_1 -(\mu_1+\mu_2),\,  p_{12} = \mu_2 -\lambda_2
\label{candidate2}
\end{split}\end{equation}
when $\lambda_{2}<\mu_{2}$, and finally

\begin{equation}\begin{split}
&\tilde{\rho}_{AB}^{(3)}=\left(1-p_{20}-p_{21}\right)\,|\Psi^{(3)}_{\textrm{mc}}\rangle\langle\Psi^{(3)}_{\textrm{mc}}|+p_{20}\,|20\rangle\langle 20| + p_{21}\,|21\rangle\langle 21|\\
&p_{20} = \lambda_1 + \lambda_1 -(\mu_1+\mu_2),\,  p_{21} = \mu_1 -\lambda_1
\label{candidate3}
\end{split}\end{equation}
when $\lambda_{1}<\mu_{1}$. As an example the matrix form of $\tilde{\rho}_{AB}^{(1)}$ reads as follows:

\begin{widetext}

\small
\begin{equation}
\setlength{\arraycolsep}{2pt}
\renewcommand{\arraystretch}{0.8}
\tilde{\rho}_{AB}^{(1)}=
\left(\begin{array}{ccc|ccc|ccc}
1-\lambda_{1}-\lambda_{2}&\cdot&\cdot&\cdot&\sqrt{\mu_{1}(1-\lambda_{1}-\lambda_{2})}&\cdot&\cdot&\cdot&\sqrt{\mu_{2}(1-\lambda_{1}-\lambda_{2})}\\
\cdot&0&\cdot&\cdot&\cdot&\cdot&\cdot&\cdot&\cdot\\
\cdot&\cdot&0&\cdot&\cdot&\cdot&\cdot&\cdot&\cdot\\
\hline
\cdot&\cdot&\cdot&\lambda_{1}-\mu_{1}&\cdot&\cdot&\cdot&\cdot&\cdot\\
\sqrt{\mu_{1}(1-\lambda_{1}-\lambda_{2})}&\cdot&\cdot&\cdot&\mu_{1}&\cdot&\cdot&\cdot&\sqrt{\mu_{1}\mu_{2}}\\
\cdot&\cdot&\cdot&\cdot&\cdot&0&\cdot&\cdot&\cdot\\
\hline
\cdot&\cdot&\cdot&\cdot&\cdot&\cdot&\lambda_{2}-\mu_{2}&\cdot&\cdot\\
\cdot&\cdot&\cdot&\cdot&\cdot&\cdot&\cdot&0&\cdot\\
\sqrt{\mu_{2}(1-\lambda_{1}-\lambda_{2})}&\cdot&\cdot&\cdot&\sqrt{\mu_{1}\mu_{2}}&\cdot&\cdot&\cdot&\mu_{2}\\
\end{array}\right).
\label{exa1}
\end{equation}
\normalsize
\end{widetext}
%
%\normalsize

The specific form of $|\Psi^{(i)}_{\textrm{mc}}\rangle\langle\Psi^{(i)}_{\textrm{mc}}|,\, i=1,2,3$ in (\ref{candidate1}, \ref{candidate2}, \ref{candidate3}) is easily found by properly adjusting partial traces.  Maximal negativity within the family (\ref{familydef}) is thus attained when $\sigma_{\textrm{mc}}$ is rank-1, that is, when the state has the following structure, similar to the two-qubit MEMMS (\ref{MEMMS}):

\begin{equation}
\tilde{\rho}_{AB}=(1-\eta)|\Psi_{\textrm{mc}}\rangle\langle\Psi_{\textrm{mc}}|+ \sum_{i\neq j} p_{ij}|ij\rangle\langle i j|
\label{struc}
\end{equation}
where $\eta=\sum_{i\neq j} p_{ij}$, namely a convex combination of a rank-1 projector and a classical state with at most two non zero entries. To conclude this section we state the following proposition (proven in Appendix A):\newline
\newline
\textbf{Proposition}: \textit{All candidate states (\ref{candidate1}, \ref{candidate2}, \ref{candidate3}) are extremal points in the convex set $C(\rho_A,\rho_B)$ of states with fixed marginals $\rho_A$ and $\rho_B$}.\newline

In what follows we show that the same states can be found from an entanglement distillation perspective, i.e.  imposing that states of the form (\ref{familydef}) are \textit{quasidistillable}.

\subsection{Quasidistillable states}

As introduced before, quasidistillable states
are mixed entangled states for which a singlet fraction arbitrarily close to unity can be distilled with non-zero
probability. In this section we recall the main feature of such states and provide a criterion to identify them within the class (\ref{familydef}). The main motivation is that the two-qubit MEMMS (\ref{MEMMS}) is also a quasidistillable state. Interestingly, we will show that all candidate states in (\ref{candidate1}, \ref{candidate2}, \ref{candidate3}) are again quasidistillable. As usual, we denote the computational basis in $\mathbb{C}^{d}\otimes\mathbb{C}^{d}$ with $\{|ij\rangle\}_{i,j=1,\dots,d}$.  Let us start from the following \cite{quasi}:\newline
\newline
\textbf{Definition}: \textit{A state $\rho$ is said \textbf{quasidistillable} iff there exist two sequences of filtering operators $\{A_n\}$ and $\{B_n\}$ such that}:

\begin{equation}
\frac{\Lambda^{(n)}(\rho)}{\textrm{Tr}\left[\Lambda^{(n)}(\rho)\right]}=\frac{(A_n\otimes B_n)\rho (A^{\dag}_n\otimes B^{\dag}_n)}{\textrm{Tr}\left[(A_n\otimes B_n) \rho (A^{\dag}_n\otimes B^{\dag}_n)\right]}  \xrightarrow[n\to\infty]{} P^{+}_{d}
\label{defdist}
\end{equation}
\textit{and the probabilities $p_n=\mathrm{Tr}[\Lambda^{(n)}(\rho)]\rightarrow 0$.}\newline

Note that the filtering operators can be taken Hermitian so we can simply restrict to $A_n$ and $B_n$. In order to characterize quasidistillable states within (\ref{familydef}), we state our two main results concerning first maximally correlated states only and the structure of our candidate MEMS with respect to fixed marginals (\ref{struc}), proven in Appendices B and C: \newline

\textbf{Theorem 1}: \textit{A maximally correlated state $\sigma_{\mathrm{mc}}$ is quasidistillable iff it is of rank 1, i.e. $\sigma_{\mathrm{mc}}=|\Psi_{\mathrm{mc}}\rangle\langle\Psi_{\mathrm{mc}}|$,  and $\mathrm{s-rank}(|\Psi_{k}\rangle)=d$ (Schmidt rank).}
\newline

\textbf{Theorem 2}: \textit{Let $\rho$ be a state of the form (\ref{struc}), i.e. a convex mixture of a maximally correlated rank-1 projector and a classical state. $\rho$ is quasidistillable iff among the set of $p_{ij}\neq 0$ there are no looping indices, i.e. $p_{ij}p_{jk}\dots p_{li}=0$. }\newline

In \cite{quasi}, the authors proved that the following two-qutrit state:

\begin{equation}
\rho = \eta P_3^+ + \frac{(1-\eta)}{3}\left(|01\rangle\langle 01|+|12\rangle\langle 12|+|20\rangle\langle 20|\right)
\label{examplequasi}
\end{equation}
with $0<\eta<1$ is not quasidistillable.  Indeed, one has $p_{01} p_{12} p_{20} \neq 0$, that is $p_{ij}$ meet the loop condition.  However, the following state
\begin{equation}
\rho = \eta P_3^+ + \frac{(1-\eta)}{3}\left(|10\rangle\langle 10|+|12\rangle\langle 12|+|20\rangle\langle 20|\right)
\end{equation}
is quasidistillable according to the sequence of filtering operators $\{A_n\}$ and $\{B_n\}$ provided in the proof of Theorem 2 in Appendix B. Furthermore, structures similar to our candidate states (\ref{candidate1}, \ref{candidate2}, \ref{candidate3}) can be recovered by means of the following:
\newline

\textbf{Corollary}: \textit{If $\rho$ of the form (\ref{struc}) is quasidistillable, it has at most  $\binom{d}{2}$ non-zero diagonal elements}. \newline

\textit{Proof}. Let $\rho$ be of the form (\ref{form}) with $p_{ij} >0 \: \forall i>j$. It easy to see from Theorem 2 that $\rho$ is quasidistillable and has exactly $\binom{d}{2}$ non-zero elements. If we consider a further non-zero element from the remaining set ($i<j$) we would have $p_{i_{0},j_{0}}p_{j_{0},i_{0}}\neq 0$ for at least one couple of indexes $(i_{0},j_{0})$ meaning that such a $\rho$ is no more quasidistillable. \newline

Therefore, only one element is allowed in the two-qubits case and at most three for two-qutrits. Some special cases  of (\ref{struc}) are the following:

\begin{equation}\begin{split}
\rho=(1-\eta)|\Psi_{\textrm{mc}}\rangle\langle\Psi_{\textrm{mc}}|+ |i_0\rangle\langle i_0|\otimes\sum^{d-1}_{j}p_{i_0,j}|j\rangle\langle j|\\
\rho=(1-\eta)|\Psi_{\textrm{mc}}\rangle\langle\Psi_{\textrm{mc}}|+ \sum^{d-1}_{i}p_{i,j_0}|i\rangle\langle i|\otimes |j_0\rangle\langle j_0|
\end{split}\end{equation}
that is, with some fixed index $i_0$ or $j_0$ in one of the two marginal subspaces.
As a final remark, we have observed that the maximization of negativity within the family (\ref{familydef}) with fixed marginals yields candidate states satisfying Theorem 2. In particular, the requirement of having the maximal number (three) of negative eigenvalues of $\rho^{\tau}$  yields at most three non-zero elements in the classical term. Moreover, the two qutrit candidates display \textit{only} two non zero $p_{i,j}$  such that the indices do not \textit{loop}, in the above sense. This leads us to conjecture that all MEMS with respect to fixed marginals are quasidistillable in arbitrary dimensions.

\section{\label{sec:level4}Numerical Results}

The aim of this section is to provide a set of numerical observations in order to legitimate our states (\ref{candidate1}, \ref{candidate2}, \ref{candidate3}) as good candidates for two-qutrits MEMS with respect to fixed marginals.
To begin with, we observe that a key ingredient is generation of random states with fixed marginals $\rho_A$ and $\rho_B$, i.e. an element of $C(\rho_A,\rho_B)$. To this aim, we have adopted two procedures. Firstly, for the two-qubits case we algorithmically generated random correlation elements of eq. (\ref{form}) and check the positivity of the resulting $\rho_{AB}$. This procedure was used to generate points in the N-P plot in fig. (\ref{random}).  A more efficient method is to choose a state randomly and to minimise numerically\footnote{We use for it the SciPy function \textit{minimize}} a distance function from the set $C(\rho_A,\rho_B)$. Such a distance is simply defined as:

\begin{equation}
\rho \mapsto || \textrm{Tr}_B \rho - \rho_A ||_2^2 + || \textrm{Tr}_A \rho - \rho_B ||_2^2.
\label{constraint}
\end{equation}

Having a random initial state from the set $C(\rho_A,\rho_B)$, we proceed maximising the negativity function. We stay in the set $C(\rho_A,\rho_B)$ during the minimization, adding the mentioned function (\ref{constraint}) to the (negated) negativity as a penalty function, with a factor controlling the accuracy.

\begin{figure}%[h!]
\hspace*{-0.55cm}\includegraphics[scale=0.28]{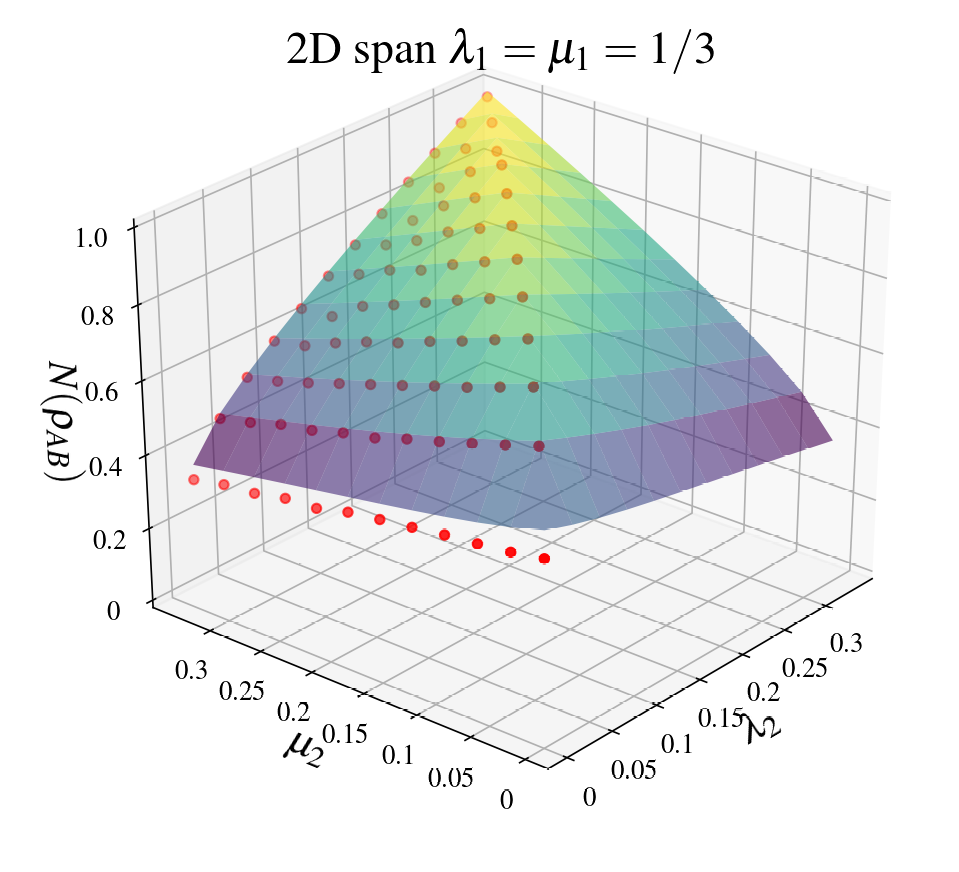}
\caption{3D plot of negativity as a function of $\lambda_2,\mu_2$. Red points obtained from numerical optimization are compared with the negativity surface obtained from our candidate.}
\label{2dspan1}
\end{figure}

In the minimization procedure, we represent states as: $\rho = A A^\dagger$, where $A$ is a square complex matrix ($9 \times 9$ for qutrits), if $\rho$ has a non restricted rank. Note, however, that according to \cite{parthasarathy} we have that $\textrm{rank}(\rho)\leq\sqrt{2d^2-1}$ for extremal states. For the two-qutrits case, the latter is $\sqrt{17}\approx 4.12$ and we can limit our search to rank-4 states only, represented by complex matrices  $A$ of size $9 \times 4$ which reduces the (real) dimension of the problem from 162 to 72.
%As an example, we chose the marginals such that $\lambda_{1}=0.25$, $\lambda_{2}=0.125$ and $\mu_{1}=0.3$, $\mu_{2}=0.1$ and show the results as matrix plots in fig. (\ref{matrixplot}). In such plots resulting from numerical optimization one may recognize the structure of a rank-1 projector and the presence of two non-zero diagonal elements according to $\tilde{\rho}^{(1)}_{AB}$ in (\ref{exa1}).
%\begin{figure}\begin{center}
%\hspace*{-0.5cm}\includegraphics[scale=0.34]{matrix.png}
%\caption{Matrix plots of the moduli of density matrix elements. The optimization in $C(\rho_A,\rho_B)$ starts from a randomly generated state (a). As a result, one can see that the negativity and purity maximization (b, c) show compatible outcomes with a particular choice of marginals. }
%\label{matrixplot}
%\end{center}
%\end{figure}
\begin{figure}\begin{center}
\hspace*{-0.3cm}\includegraphics[scale=0.3]{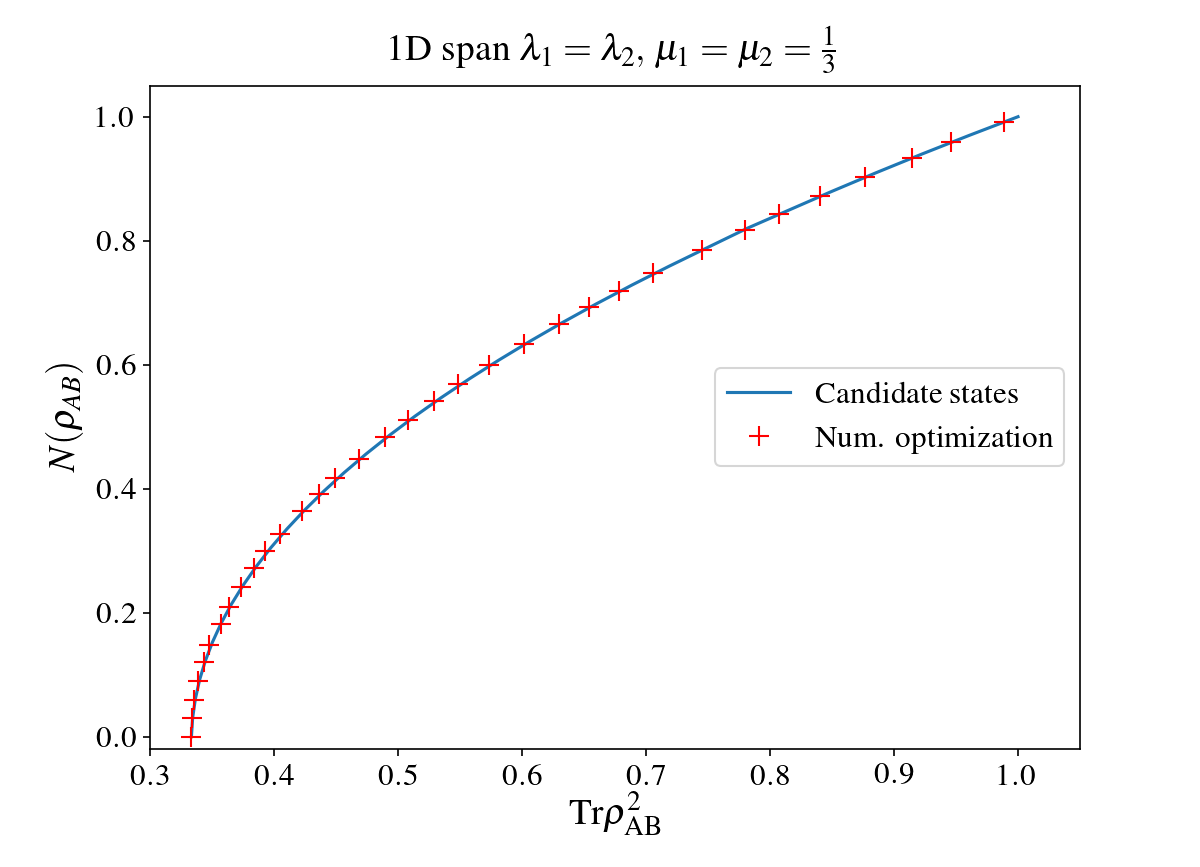}
\caption{Negativity vs. global purity (N-P) plot. We analyze a particular configuration with one marginal maximally mixed and another spanning over one eigenvalue only, namely  $\lambda_1=\lambda_2$. The blue line describes the negativity of the candidate states (\ref{candidate1}) for $P\in[\frac{1}{3},1]$. Red crosses represent the negativity values obtained from numerical optimization.}
\label{1dspan}
\end{center}
\end{figure}
A restricted, one-dimensional set of examples is shown in fig. (\ref{1dspan}) where one can see a satisfactory agreement between the negativity of the candidate states (blue line) and the results of numerical optimization (red crosses) for a particular set of marginals.
A second set of examples is obtained spanning over the two lowest marginal eigenvalues independently, thus keeping $\lambda_1$ and $\mu_1$ fixed. For the set of points in fig. (\ref{2dspan1}) we choose $\lambda_1 = \mu_1 = \frac{1}{3}$ and span over uniformly distributed values of the allowed domain for $\lambda_2$ and $\mu_2$ and compare the negativity surface from the candidate states with numerical optimums.  Note that for such choice we have one candidate only since all candidate states collapse in one.
\begin{figure}%[h!]
\hspace*{-0.55cm}\includegraphics[scale=0.28]{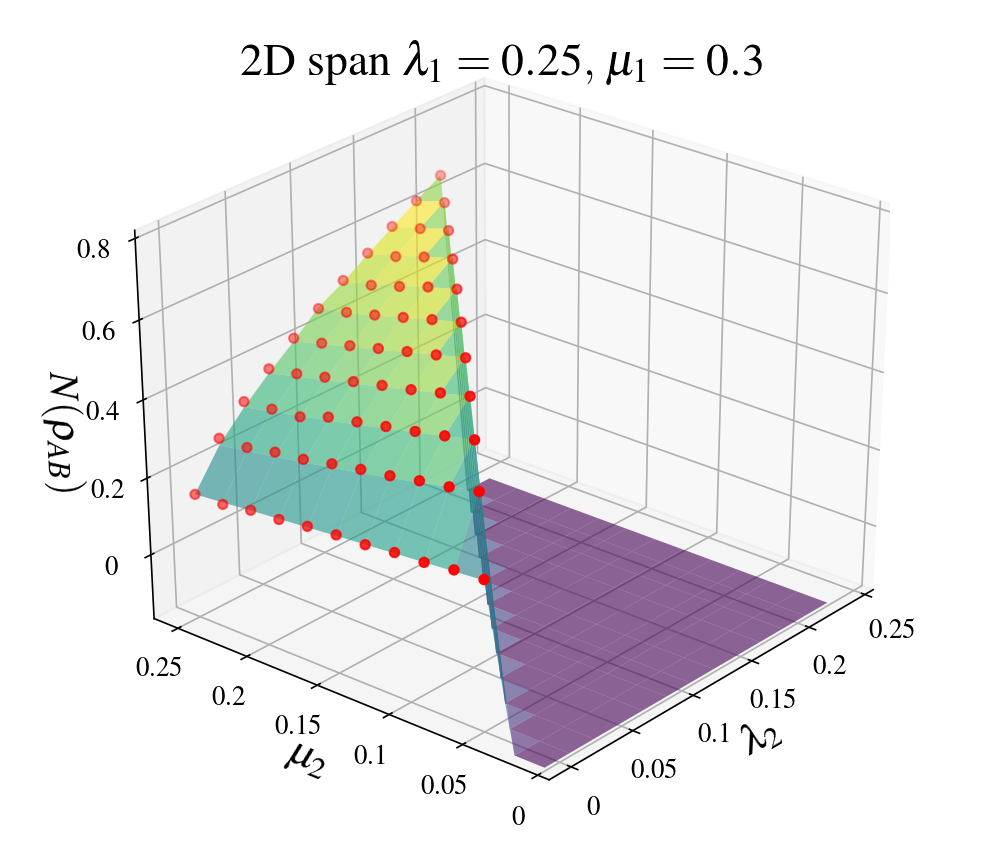}
\caption{Second 3D plot of negativity as a function of the lowest marginal eigenvalues $\lambda_2,\mu_2$. Here $\lambda_1,\mu_1$ are chosen such that the domain of interest is restricted.}
\label{2dspan2}
\end{figure}
As a last series of examples we choose $\lambda_1 = 0.25,\, \mu_1 = 0.3$ so that the candidate is given by $\tilde{\rho}^{(3)}_{AB}$  in (\ref{candidate3}) and the range for $\lambda_2,\mu_2$ is restricted by the assumption $\lambda_{1}+\lambda_{2}\geq\mu_{1}+\mu_{2}$ (See fig. (\ref{2dspan2})). To summarize, all the above results strongly support our conjecture that our quasidistillable states (\ref{candidate1}, \ref{candidate2}, \ref{candidate3}) are legitimate candidates for two-qutrits MEMS with respect to fixed marginals and motivate the search for analytical proofs in further studies.

\section{Conclusive Remarks}

In this work we have observed a strong numerical evidence that the states (\ref{candidate1}, \ref{candidate2}, \ref{candidate3}) are indeed good candidates as MEMS with respect to fixed marginals. The main feature of our reasoning is the generalization of two special properties of the two-qubit state (\ref{MEMMS}), i.e. its simple structure and the property of being quasidistillable. It is shown that these states are always  quasidistillable and hence we provide another interesting application of quasidistillable states in quantum information. Such a strong link between the two concepts deserves to be investigated in further studies. Moreover, a possible obvious generalization of our problem can be thought for multipartite entanglement in the presence of many fixed marginal states. Other similar versions can be considered such as the bounds of mutual information, coherence or the study of the such bounds in the presence of fixed marginal purities, as the original problem in \cite{adesso}. The difference is the corresponding set of states is not convex and we cannot rely on the extremality property.
Finally, concerning our problem, it is worth remarking the that both the maximization of negativity and purity lead to the same optimal state. This is true for the two-qubit case and for the two-qutrit family defined by (\ref{familydef}) and there is numerical evidence for general two-qutrit states. This observation will be also object of further investigations.
We hope that further characterizations of extremal points in $C(\rho_A,\rho_B)$ in future studies could lead to other observations strengthening our conjecture and pave the way to analytical proofs.

\section*{Acknowledgements}

GB acknowledges R. Ayllon and R. Palacino for useful discussions. DC and GS were supported by the Polish National Science Centre project 2015/19/B/ST1/03095. PH acknowledges support by the Foundation for Polish Science through IRAP project co-financed by EU within
Smart Growth Operational Programme (contract no. 2018/MAB/5). 

\section*{Appendix A (Extremal states)}

In this appendix we show that the two-qubit MEMMS (\ref{MEMMS}) and all the candidate states (\ref{candidate1}, \ref{candidate2}, \ref{candidate3}) are extremal points in the convex set of states with fixed marginals $C(\rho_A,\rho_B)$. According to the conditions (\ref{extcond1}) and (\ref{extcond2}), a generic state in $\mathbb{C}^d\otimes\mathbb{C}^d$, $d\geq2$, defined as:

\begin{equation}
\rho_{\Lambda} = [\textrm{id}_{d}\otimes\Lambda](P^+_{d})=\rho_{\Lambda}=\frac{1}{d}\sum^{d}_{i,j=1}\sum^{d^2}_{\alpha=1}e_{ij}\otimes K_{\alpha}\,e_{ij}\,K^{\dagger}_\alpha
\end{equation}

is extremal in $C(\rho_A,\rho_B)$ iff $\Lambda(\mathbb{I}_{d})=d\rho_{A}$, $\Lambda^*(\mathbb{I}_{d})=d\rho_{B}$ and the set $\{K^{\dagger}_{\alpha}K_{\beta}\oplus K_{\beta}K^{\dagger}_{\alpha}\}_{\alpha,\beta=1,\dots,d^2}$ is linearly independent. For the rank-2 two-qubit MEMMS state (\ref{MEMMS}), we have the following suitable family of Kraus operators:

\begin{equation}\begin{split}
K_{1} = \left(\begin{array}{cc}
0 & a \\
b & 0
\end{array}\right),\quad K_{2} = \left(\begin{array}{cc}
x & 0 \\
0 & y
\end{array}\right)\\
K_{1}K^{\dagger}_{1} + K_{2}K^{\dagger}_{2} = \left(
\begin{array}{cc}
 |a|^2+|x|^2 & 0 \\
 0 & |b|^2+|y|^2 \\
\end{array}
\right)\\
K^{\dagger}_{1}K_{1} + K^{\dagger}_{2}K_{2} = \left(
\begin{array}{cc}
 |b|^2+|x|^2 & 0 \\
 0 & |a|^2+|y|^2 \\
\end{array}
\right).
\label{krausapp}
\end{split}\end{equation}

Choosing $a=0$ implies $|x|=\sqrt{1-\lambda_{A}}$, $|y|=\sqrt{\lambda_{B}}$ and $|b|=\sqrt{\lambda_A-\lambda_B}$. Moreover the Kraus operators satisfy:

\begin{equation}\begin{split}
K^{\dagger}_{1}K_{2}=\sqrt{\lambda_{B}(\lambda_{A}-\lambda_{B})}|0\rangle\langle 1| = (K^{\dagger}_{2}K_{1})^{\dagger},\\
K_{1}K^{\dagger}_{2}=\sqrt{(1-\lambda_{B})(\lambda_{A}-\lambda_{B})}|1\rangle\langle 0| = (K_{2}K^{\dagger}_{1})^{\dagger}.
\end{split}\end{equation}

Thus, the two sets $\{K^{\dagger}_{\alpha}K_{\beta}\}_{\alpha,\beta=1,2}$ and $\{K_{\beta}K^{\dagger}_{\alpha}\}_{\alpha,\beta=1,2}$ are jointly linear independent and we have:

\begin{equation*}
\rho_{\Lambda}=\tilde{\rho}_{AB}=\left(\begin{array}{cccc}
1-\lambda_{A}&\cdot&\cdot&\sqrt{(1-\lambda_{A})\lambda_{B}}\\
\cdot&0&\cdot&\cdot\\
\cdot&\cdot&\lambda_{A}-\lambda_{B}&\cdot\\
\sqrt{(1-\lambda_{A})\lambda_{B}}&\cdot&\cdot&\lambda_{B}\end{array}\right).
\end{equation*}

By means of a similar argument one finds the corresponding Kraus operators for the candidate states $\tilde{\rho}^{(i)}_{AB}, \, i=1,2,3$. We have for $\tilde{\rho}^{(1)}_{AB}$:

\begin{equation*}\begin{split}
&K_{1} = \sqrt{1-\lambda_{1}-\lambda_{2}}\,|0\rangle\langle 0| + \sqrt{\mu_1}\,|1\rangle\langle 1|+\sqrt{\mu_2}\,|2\rangle\langle 2|\\
&K_{2} = \sqrt{\lambda_1-\mu_1}\,|1\rangle\langle 0|,\quad K_{3} = \sqrt{\lambda_2-\mu_2}\,|2\rangle \langle 0|
\end{split}\end{equation*}

which produce the following state via (\ref{krausapp}):

\small
\begin{equation*}
\rho_{\Lambda}=\left(\begin{array}{ccc|ccc|ccc}
\alpha^{2}_{00}&\cdot&\cdot&\cdot&\alpha_{00}\alpha_{11}&\cdot&\cdot&\cdot&\alpha_{00}\alpha_{22}\\
 &0&\cdot&\cdot&\cdot&\cdot&\cdot&\cdot&\cdot\\
 & &0&\cdot&\cdot&\cdot&\cdot&\cdot&\cdot\\
\hline
 &&&\lambda_{1}-\mu_{1}&\cdot&\cdot&\cdot&\cdot&\cdot\\
 &&&&\alpha^{2}_{11}&\cdot&\cdot&\cdot&\alpha_{11}\alpha_{22}\\
 &&&&&0&\cdot&\cdot&\cdot\\
\hline
 &&&&&&\lambda_{2}-\mu_{2}&\cdot&\cdot\\
 &&&&&&&0&\cdot\\
 &&&&&&&&\mu_{2}\\
\end{array}\right)
\end{equation*}
\normalsize

where $\alpha_{00}= \sqrt{1-\lambda_{1}-\lambda_{2}}$, $\alpha_{11} = \sqrt{\mu_1} $,  $\alpha_{22} = \sqrt{\mu_2} $. One sees that such a state coincides with $\tilde{\rho}^{(1)}_{AB}$ (\ref{exa1}). Other possible Kraus operators for the states (\ref{candidate2}, \ref{candidate3}) are found as:

\begin{equation*}\begin{split}
&K_{1} = \sqrt{1-\lambda_{1}-\lambda_{2}}\,|0\rangle\langle 0| + \sqrt{\mu_1}\,|1\rangle\langle 1|+\sqrt{\lambda_2}\,|2\rangle\langle 2|\\
&K_{2} = \sqrt{\lambda_1+\lambda_2-(\mu_1+\mu_2)}\,|1\rangle\langle 0|,\quad K_{3} = \sqrt{\mu_{2}-\lambda_2}\,|1\rangle \langle 2|,
\end{split}\end{equation*}

\begin{equation*}\begin{split}
&K_{1} = \sqrt{1-\lambda_{1}-\lambda_{2}}\,|0\rangle\langle 0| + \sqrt{\lambda_1}\,|1\rangle\langle 1|+\sqrt{\mu_2}\,|2\rangle\langle 2|\\
&K_{2} = \sqrt{\mu_2-\lambda_2}\,|2\rangle\langle 1|,\quad K_{3} = \sqrt{\lambda_1+\lambda_2-(\mu_1+\mu_2)}\,|2\rangle \langle 0|
\end{split}\end{equation*}

valid for $\tilde{\rho}^{(2)}_{AB}$ and $\tilde{\rho}^{(3)}_{AB}$ respectively.

\section*{Appendix B (Theorem 1)}

Before proving Theorem 1 let us state the following lemma concerning filtering operators $A^T_n$ and $B_n$.\newline

\textbf{Lemma 1}: \textit{Let $\{A^T_n\}$ and $\{B_n\}$ be filtering operators for some state $\rho$ in quasidistillation process and $0\leq a^{(n)}_1\leq \dots\leq a^{(n)}_d$ and $0\leq b^{(n)}_1\leq \dots\leq b^{(n)}_d$ their singular eigenvalues. Then, at least one among $a^{(n)}_1, b^{(n)}_1$ must tend to zero as $n\rightarrow\infty$.}\newline

\textit{Proof}. Suppose that both $a^{(n)}_1, b^{(n)}_1\geq \gamma >0\quad \forall n$ and let us consider $|\Psi\rangle=\sum_{i}d_{ii}|ii\rangle$ satisfying eq. (\ref{defdist}). Then $\mathrm{s-rank}(|\Psi_{k}\rangle)=d$ and the  matrix $D=\{d_{ii}\}$ has full rank, i.e. $d_{ii}\geq\delta > 0$. Eq. (\ref{defdist}) is then equivalent to:

\begin{equation}
\frac{B_n D A_n}{\|B_n D A_n\|_{\textrm{H-S}}} \xrightarrow[n\to\infty]{} \frac{\mathbb{I}}{\sqrt{d}},
\label{xn}
\end{equation}
where $\|\omega\|_{\textrm{H-S}}=\sqrt{{\rm Tr}\omega\omega^{\dag}}$ is the Hilbert-Schmidt norm. Note that $\|B_n D A_n\|_{\textrm{H-S}}=\textrm{Tr}\left[\Lambda^{(n)}(|\Psi\rangle\langle\Psi|)\right]^{\frac{1}{2}}$ so that it must tend to zero as $n\rightarrow \infty$. However, we have the following:

\begin{equation*}\begin{split}
&\textrm{Tr}\left[\Lambda^{(n)}(|\Psi\rangle\langle\Psi|)\right]=\textrm{Tr}\left[B_n D A_n A_n D B_n\right]\geq \\
&\textrm{Tr}\left[B_n D^{2} B_n\right]\gamma^{2}\geq \textrm{Tr}\left[A^{2}\right]\gamma^{2}\delta \geq \gamma^{4}\delta > 0.
\end{split}\end{equation*}
Therefore, at least one among $a^{(n)}_1, b^{(n)}_1$ must tend to zero. Let us now prove Theorem 1. \newline

\textit{Proof $\,\,$(Theorem 1)}. Consider  the quasidistillation of $\sigma_{\textrm{mc}}$ (\ref{familydef})
which has many eigenvectors $|\Psi_k \rangle$. As  already mentioned, they all have diagonal coefficient matrices  $D^{k}$ with elements
$D^{k}_{ij}=\lambda_{ij}^{(k)}$. Because of quasidistillation process, at least one of the eigenvectors satisfies (in terms of $|\Psi_k \rangle\langle \Psi_k|$) eq. (\ref{defdist}) which we shall drop a particular index denoting  that vector and its coefficients matrix as $|\Psi \rangle$ and $D$ accordingly. We shall show that if the mixture $\sigma_{\textrm{mc}}$ is to satisfy (\ref{defdist}) then it cannot admit any more eigenvectors but $|\Psi \rangle $.
\linebreak
	
Let  $|\Psi ' \rangle = \sum_{i} (d')_{ii} |i i\rangle $ be another arbitrary eigenvector with its corresponding coefficients matrix $D'$.
We shall show that either it vanishes or is proportional to $|\Psi\rangle$.
There are three alternatives: the ratio	$\frac{\Lambda^{(n)}(|\Psi '  \rangle \langle \Psi ' | )}{|\Psi\rangle \langle \Psi |}$
can (i) converge to a strictly positive constant (ii), diverge to infinity or (iii) converge to zero\footnote{If there are some oscillations in those sequences, then we can always find  subsequences of filters that realize quasidistillation, satisfying the classification (i-iii).}.
This corresponds to the situations that a weight at the transformed
eigenvector $|\Psi \rangle $ is comparable,  dominates or is
dominated in the limit of large $n$ respectively. 

Consider first the case (i). Here we have:

\begin{equation*}
\frac{\sqrt{\textrm{Tr}\left[\Lambda^{(n)}(|\Psi\rangle\langle\Psi|)\right]}}{\sqrt{\textrm{Tr}\left[\Lambda^{(n)}(|\Psi'\rangle\langle\Psi'|)\right]}} =\frac{\|B_n D A_n\|_{\textrm{H-S}}}{\|B_n D' A_n\|_{\textrm{H-S}}}\xrightarrow[n\to\infty]{}c>0,
\end{equation*}
where of course $D'= \{d'_{ii}\} $. If we call $X_n$ the LHS of eq. (\ref{xn}), we have by assumption the following:

\begin{equation} \label{matrixlimit}
X'_n=\frac{B_n D' A_n}{\|B_n D' A_n\|_{\textrm{H-S}}}\xrightarrow[n\to\infty]{} \frac{\mathbb{I}}{\sqrt{d}}.
\end{equation}
Both $X_n$ and $X'_n$ have bounded inversion so we have:

\begin{equation}\begin{split}
&X_n\left(X'_n\right)^{-1} =\\
&\frac{\|B_n D' A_n\|_{\textrm{H-S}}}{\|B_n D A_n\|_{\textrm{H-S}}} B_n D A_n (A_n)^{-1}(D')^{-1}(B_n)^{-1} \xrightarrow[n\to\infty]{} \mathbb{I},
\end{split}\end{equation}
or, equivalently:

\begin{equation}
B_n D (D')^{-1} (B_n)^{-1} \xrightarrow[n\to\infty]{} c \mathbb{I}.
\label{oper}
\end{equation}
Let us now transpose eq. (\ref{oper}) into its matrix representation in the basis $\left\{|b^{(n)}_i\rangle\right\}$ of the eigenvectors of $B_n$ corresponding to the increasingly ordered eigenvalues $b^{(n)}_i$:

\begin{equation}\begin{split}
&\langle b^{(n)}_i|B_n D (D')^{-1} (B_n)^{-1}|b^{(n)}_j\rangle =\\
& b^{(n)}_i (b^{(n)}_j)^{-1} \langle b^{(n)}_i| D (D')^{-1}|b^{(n)}_j\rangle \xrightarrow[n\to\infty]{} c\delta_{ij}.
\label{limit}
\end{split}\end{equation}
The products $b^{(n)}_i (b^{(n)}_j)^{-1}$ define a set of coefficients which can be represented in the following matrix form:

\begin{equation}\begin{split}
\left(\begin{array}{c}
b^{(n)}_1\\
\vdots\\
b^{(n)}_d
\end{array}\right)\cdot\left(\begin{array}{cccc}
(b^{(n)}_1)^{-1}&\dots&(b^{(n)}_n)^{-1}\end{array}\right)=
&\\
\left(\begin{array}{cccc}
1&b^{(n)}_1(b^{(n)}_2)^{-1}&b^{(n)}_1(b^{(n)}_3)^{-1}&\dots\\
b^{(n)}_2(b^{(n)}_1)^{-1}&1&b^{(n)}_2(b^{(n)}_3)^{-1}&\dots\\
b^{(n)}_3(b^{(n)}_1)^{-1}&b^{(n)}_3(b^{(n)}_2)^{-1}&1&\\
\vdots&\vdots&&\ddots
\end{array}\right),
\end{split}
\label{matrix-of-b}
\end{equation}
in which we can easily see that each element $b^{(n)}_i (b^{(n)}_j)^{-1}\geq 1$ in the lower triangle. Therefore, in order to have eq. (\ref{limit}) satisfied, $D(D')^{-1}$ must be upper triangular in the basis of the eigenvectors of $B_n$.
{However, since $D$ and $D'$ commute,  its product is hermitian so it must be
such in the basis $|b_i \rangle $ being the limit
of the eigenbases $|b^{(n)}_i\rangle$
(again, in a sense of compactness argument).
This means eventually that it must be diagonal in that limit, which
leads to the conclusion
%However, $D(D')^{-1}$ must be diagonal in the limit basis and therefore we have
 that $\langle b_i| D (D')^{-1}|b_j\rangle = c \delta_{ij}$
 %$ \:\forall n$.
 In other words, $D \propto D'$, so effectively
 $|\Psi'\rangle$ is proportional to $|\Psi\rangle$ and in this sense
 removed form the eigenrepresentation of $\sigma_{\textrm{mc}}$ . }

%Let us suppose now $\sigma_{\textrm{MC}}=p_{1}|\Psi\rangle\langle\Psi| +p_{2}|\Psi'\rangle\langle\Psi'|$ with:
Consider now the case (ii) from the alternative options (i-iii).
Here we have by assumption:

\begin{equation}
\frac{\textrm{Tr}\left[\Lambda^{(n)}(|\Psi'\rangle\langle\Psi'|)\right]}{\textrm{Tr}\left[\Lambda^{(n)}(|\Psi\rangle\langle\Psi|)\right]}=\frac{\|B_n D' A_n\|_{\textrm{H-S}}}{\|B_n D A_n\|_{\textrm{H-S}}}\xrightarrow[n\to\infty]{} 0.
\label{norms-convergence}
\end{equation}
Therefore, eq. (\ref{matrixlimit}) becomes:

\begin{equation*}
X'_n=\frac{B_n D' A_n}{\|B_n D' A_n\|_{\textrm{H-S}}}\xrightarrow[n\to\infty]{} \mathbf{0}.
\end{equation*}
Let us consider this time the product $X'_n(X_n)^{-1}$:

\begin{equation}\begin{split}
&X'_n(X_n)^{-1}=\\
& \frac{\|B_n D A_n\|_{\textrm{H-S}}}{\|B_n D' A_n\|_{\textrm{H-S}}} B_n D' A_n (A_n)^{-1}D^{-1}(B_n)^{-1}\xrightarrow[n\to\infty]{} \mathbf{0},
\label{x'x}
\end{split}\end{equation}
that, applying the same above reasoning, it becomes:

\begin{equation*}
\frac{\|B_n D A_n\|_{\textrm{H-S}}}{\|B_n D' A_n\|_{\textrm{H-S}}}\cdot b^{(n)}_i (b^{(n)}_j)^{-1}\cdot \langle b^{(n)}_i| D 'D^{-1}|b^{(n)}_j\rangle \xrightarrow[n\to\infty]{} 0.
\end{equation*}

{Assumption (\ref{norms-convergence}) implies that
the fraction of norms diverges in the above formula. Thus, again by the property of the matrix (\ref{matrix-of-b}), we have  that the matrix $D (D')^{-1}$ must be \textit{strictly} upper triangular (i.e. with vanishing diagonal) in the limit basis, which, by its hermiticity, implies
that
%(\ref{x'x})
$\langle b_i| D 'D^{-1}|b_j\rangle = 0 $.
Thus, since $D$ is invertible, $D'=\mathbf{0}$ which means
that $|\Psi'\rangle$ compatible with (ii) cannot exist. The last case (iii) can be immediately resolved by permuting
the roles of $|\Psi \rangle $ and $|\Psi' \rangle $
and concluding that $|\Psi\rangle$ cannot vanish by assumption
which leads to the expected contradiction.}

\section*{Appendix C (Theorem 2)}

\textit{Proof}. Since the sum in eq. (\ref{struc}) is separable, it must tend to zero when applying filtering, namely:

\begin{equation*}
\frac{1}{\textrm{Tr}\left[\Lambda^{(n)}(\rho)\right]}\sum_{i\neq j}p_{ij}\Lambda^{(n)}(|ij\rangle\langle i j|)\xrightarrow[n\to\infty]{} 0.
\end{equation*}

Moreover, due to Theorem 1, we also have that quasidistillability implies that each eigenvector must vanish in the limit when applying filtering:

\begin{equation}
\frac{\Lambda^{(n)}(|ij\rangle\langle ij|)}{\textrm{Tr}\left[\Lambda^{(n)}(|ij\rangle\langle ij|)\right]} \xrightarrow[n\to\infty]{} 0\quad i\neq j.
\label{limitij}
\end{equation}

Let us then apply $\Lambda^{(n)}$ to a generic state $|\Psi\rangle=\sum_{i,j=1}\alpha_{ij}|ij\rangle$. It is easy to see that the state:

\begin{equation*}\begin{split}
&\sum_{i,j}\alpha_{ij}\!\left(\frac{A_{n}|i\rangle}{\textrm{Tr}\left[\Lambda^{(n)}(|\tilde{\Psi}\rangle\langle\tilde{\Psi}|)^{1/4}\right]}\right)\!\otimes\!\left(\frac{B_{n}|j\rangle}{\textrm{Tr}\left[\Lambda^{(n)}(|\tilde{\Psi}\rangle\langle\tilde{\Psi}|)^{1/4}\right]}\right)\\
&=\sum_{i,j}\alpha_{ij}|a^{(i)}_{n}, b^{(j)}_{n}\rangle
\end{split}\end{equation*}

is normalized and that Eq. (\ref{limitij}) is then equivalent to $\|a^{(i)}_{n}\|\cdot\|b^{(j)}_{n}\|\rightarrow 0\:\:\: \forall i\neq j$. and thus:

\begin{equation*}
\prod_{i\neq j}\|a^{(i)}_{n}\|\cdot\|b^{(j)}_{n}\|\xrightarrow[n\to\infty]{} 0
\end{equation*}

Therefore, if there is a loop in the set of indexes (i.e. $p_{ij}p_{jk}\dots p_{li} \neq 0$) we have:

\begin{equation*}
\|a^{(i)}_{n}\|\cdot\|b^{(j)}_{n}\|\cdot\|a^{(j)}_{n}\|\cdot\|b^{(k)}_{n}\|\cdot \dots \cdot \|a^{(l)}_{n}\|\cdot\|b^{(i)}_{n}\|   \xrightarrow[n\to\infty]{} 0
\end{equation*}

which after suitable reordering gives:

\begin{equation}\begin{split}
&\left(\|a^{(i)}_{n}\|\cdot\|b^{(i)}_{n}\|\right)\left(\|a^{(j)}_{n}\|\cdot\|b^{(j)}_{n}\|\right) \dots \\
&\dots\left( \|a^{(l)}_{n}\|\cdot\|b^{(l)}_{n}\| \right) \xrightarrow[n\to\infty]{} 0.
\label{reordering}
\end{split}\end{equation}

Eq. (\ref{reordering}) implies that at least one among $\left(\|a^{(i)}_{n}\|\cdot\|b^{(i)}_{n}\|\right)$ would vanish in the limit and thus the maximally correlated part $\sigma_{\textrm{MC}}$ cannot have maximal Schmidt rank. This argument proves that if $\rho$ has the form (\ref{struc}) and it is quasidistillable, then necessarily $p_{ij}p_{jk}\dots p_{li}=0$. In what follows, we show that this condition is also sufficient for quasidistillability.

Let $A_{n}$ and $B_n$ be operators with the following representation in the computational basis:

\begin{equation*}
A_n=\left(\begin{array}{ccccc}
n^{\alpha_1 -1}&0&0&\cdots&0\\
0&n^{\alpha_2 -1}&0&\cdots&0\\
\vdots&&\ddots&&\vdots\\
\vdots&&&\ddots&0\\
0&\cdots&\cdots&0&n^{\alpha_{d}-1}
\end{array}
\right),
\end{equation*}

\begin{equation*}
 B_n=\left(\begin{array}{ccccc}
n^{-\alpha_1}&0&0&\cdots&0\\
0&n^{-\alpha_2}&0&\cdots&0\\
\vdots&&\ddots&&\vdots\\
\vdots&&&\ddots&0\\
0&\cdots&\cdots&0&n^{-\alpha_{d}}
\end{array}
\right)
\end{equation*}

Where $\alpha_i \in [0,\frac{1}{2}]$ is a set of real numbers. Note that the stucture of $A_n$ and $B_n$ is in accordance with the result proved in Lemma 1 since, in particular, all eigenvalues vanish in the limit.
One can easily see that the filtering map $\Lambda^{(n)}$ defined by this two operators yelds the following:

\begin{equation*}
\Lambda^{(n)}\left[|i\rangle\langle i|\otimes|i\rangle\langle i|\right]=\frac{1}{n^2}|i\rangle\langle i|\otimes|i\rangle\langle i|,
\end{equation*}

\begin{equation*}
\Lambda^{(n)}\left[|i\rangle\langle j|\otimes|i\rangle\langle j|\right]=\frac{1}{n^2}|i\rangle\langle j|\otimes|i\rangle\langle j|,
\end{equation*}

\begin{equation*}
\Lambda^{(n)}\left[|i\rangle\langle i|\otimes|j\rangle\langle j|\right]=\frac{1}{n^2}\left(|i\rangle\langle i|\otimes|j\rangle\langle j|\right)n^{2(\alpha_{i}-\alpha_{j})}.
\end{equation*}

In other words, $A_n$ and $B_n$ are constructed in such a way to distill a state $\rho$ of the form (\ref{form}) iff all the inequalities $\alpha_{i}<\alpha_{j}$ hold for every $i\neq j$. We can also see that if there are no loops of indexes the inequalities  $\alpha_{i}<\alpha_{j} \: \forall i\neq j$ amount to a certain number $p$ of order relations between at least $p+1$ real numbers. Such a set is always compatible and, therefore, it is always possible to choose $\{\alpha_{i}\}$ in such a way that $A_{n}$ and $B_{n}$ filter any $\rho$ of the form (\ref{struc}).

\end{document}